\documentclass[12pt,preprint]{aastex}

\def\h0 {$h_0$=70 km s$^{-1}$ Mpc$^{-1}$}
\def\ergs { erg s$^{-1}$}
\def\ll {$L/L_{\rm Edd}$}
\def\ga {$\Gamma_{\rm 2-10\ keV}$}

\newcommand{\be}{\begin{equation}}
\newcommand{\ee}{\end{equation}}
\newcommand{\ce}{\ifmmode {\cal E} \else ${\cal E}$\ \fi}
\newcommand{\kms}{\ifmmode {\rm km\ s}^{-1} \else km s$^{-1}$\ \fi}
\newcommand{\tes}{\ifmmode \tau {\rm es} \else $\tau {\rm es}$\ \fi}
\newcommand{\tk}{\ifmmode \tau {\rm K} \else $\tau {\rm K}$\ \fi}
\newcommand{\vfwhm}{\ifmmode V {\mbox{\tiny FWHM}} \else
            $V {\mbox{\tiny FWHM}}$\fi}
\newcommand{\msun}{\ifmmode M_{\odot} \else $M_{\odot}$\ \fi}
\newcommand{\afe}{\ifmmode {\mathcal A {\rm Fe}} \else${\mathcal A {\rm Fe}}$\ \fi}

\newcommand{\lb}{\ifmmode L_{\rm Bol} \else $L_{\rm Bol}$\ \fi}
\newcommand{\ledd}{\ifmmode L_{\rm Edd} \else $L_{\rm Edd}$\ \fi}
\newcommand{\lx}{\ifmmode L_{\rm 2-10\ keV} \else  $L_{\rm 2-10\ keV}$\ \fi}
\newcommand{\hb}{\ifmmode H\beta \else H$\beta$\ \fi}
\newcommand{\mbh}{\ifmmode M_{\rm BH}  \else $M_{\rm BH}$\ \fi}
\newcommand{\lv}{\ifmmode \lambda L_{\lambda}(5100\AA) \else $\lambda L_{\lambda}(5100\AA)$\ \fi}

\def\ka{K$\alpha$~}

\def\ariel5{{\it Ariel 5}\ }

\def\xmm{{\it XMM-Newton}\ }
\def\chandra{{\it Chandra}\ }

\def\heao1{{\it HEAO~1}\ }

\shorttitle{X-ray properties of Seyfert 1 galaxies} \shortauthors{Zhou \& Zhang}

\begin{document}
\title{A comparison of hard X-ray photon indices and iron \ka emission lines in X-ray luminous narrow- and broad-line
Seyfert 1 galaxies}

\author{Xin-Lin Zhou\altaffilmark{}}
\affil{Key Laboratory of Optical Astronomy, National Astronomical
Observatories, Chinese Academy of Sciences, Beijing 100012, China}
\affil{Department of Physics and Tsinghua Center for Astrophysics,
Tsinghua University, Beijing 100084, China}

\author{Shuang-Nan Zhang\altaffilmark{}}
\affil{Key Laboratory of Particle Astrophysics, Institute of High
Energy Physics, Chinese Academy of Sciences,, Beijing 100049, China}
\affil{Physics Department, University of Alabama in Huntsville,
Huntsville, AL 35899, USA}

\email{zhouxl@nao.cas.cn, zhangsn@ihep.ac.cn}

\begin{abstract}
We use publicly available XMM-Newton data to systematically compare
the hard X-ray photon indices, $\Gamma_{\rm 2-10\ keV}$ and the iron
K$\alpha$ emission lines of narrow-line (NL) and broad-line Seyfert
1 (BLS1) galaxies. We compile a flux-limited ($f_{\rm 2-10\ keV}
\geq 1 \times 10^{-12}$ erg s$^{-1}$ cm$^{-2}$) sample including 114
radio-quiet objects, with the $2-10$ keV luminosity ranging from
10$^{41}$ to 10$^{45}$ erg s$^{-1}$. Our main results are: 1) NLS1s
and BLS1s show similar luminosity distributions; 2) The weighted
mean of $\Gamma_{\rm 2-10\ keV}$ of NLS1s, BLS1s and the total
sample is $2.04\pm0.04$, $1.74\pm0.02$, $1.84\pm0.02$, respectively;
a significant anti-correlation between \ga~ and FWHMH$\beta$
suggests that $\Gamma_{\rm 2-10\ keV}
> 2.0$ may be taken to indicate X-ray luminous NLS1 type; 3) The 6.4 keV
narrow iron K$\alpha$ lines from NLS1s are generally weaker than
that from BLS1s; this would indicate a smaller covering factor of
the dusty tori in NLS1s, if the line emission originates from the
inner boundary region of the dusty torus in an AGN; 4) all the
broadened iron K$\alpha$ lines with intrinsic width $\sigma>0.5$ keV
correspond to FWHM\hb $\leq 4000 ~\kms$.

\end{abstract}


\section{Introduction}

It was found that a power-law component dominates the observed X-ray spectra of Seyfert 1 galaxies above 2 keV
(Mushotzky et al. 1980). Other complex spectral features can be modeled by reprocessing of the hard power-law
continuum in the circumnuclear cold matter (Pounds et al. 1990), with the mean photon index $\Gamma\sim 1.9-2.0$
(Nandra \& Pounds 1994), where $f(E) \sim E^{-\Gamma}$. Therefore, the hard photon index, \ga, characterize the
basic X-ray spectral shape of Seyfert galaxies, giving important clues to the emission mechanism and source
properties.

Narrow-line Seyfert 1 (NLS1) galaxies (Osterbrock \& Pogge 1985) are
a subset of Seyfert 1 galaxies conventionally defined from their
optical parameters (see Komossa 2008 and references therein). Many
previous studies have suggested that NLS1 galaxies may have softer
X-ray spectra than those of broad-line Seyfert 1 (BLS1) galaxies
(Laor et al. 1994; Boller, Brandt \& Fink 1996; Wang, Brinkmann \&
Bergeron 1996). There is an anti-correlation between \ga~ and FWHM
of the optical \hb lines found in Seyfert 1 galaxies (Brandt, Mathur
\& Elvis 1997). Therefore, NLS1 galaxies have been generally
believed to show softer X-ray spectra, similar to the
high/soft-state spectra from Galactic X-ray binaries (Pounds, Done
\& Osborne 1995; Zhou et al. 2007). However, recent studies
suggested that the anti-correlation shows large scatter (Piconcelli
et al. 2005) and some NLS1 galaxies selected from the Sloan Digital
Sky Survey do not display softer X-ray spectra (Zhou et al. 2006).

Fluorescent iron \ka lines are the most prominent reprocessed features in the X-ray spectra of Seyfert galaxies.
\xmm observations revealed that a narrow iron \ka line (NIKAL) at 6.4 keV is almost ubiquitous, with a
substantial fraction of objects showing broadened iron \ka line (BIKAL) emission (Reeves et al. 2006; Nandra et
al. 2007, N07). NIKAL was suggested to provide information of the geometry of the molecular torus, indicating
changing AGN populations (Zhou \& Wang 2005; Bianchi et al. 2007). BIKAL probes the strong gravity and spin of a
black hole (e.g., Fabian et al. 2009; see the review by Reynolds \& Nowak 2003 and Miller 2007). Note that the
line profile seems to be very different in different sources. Studying the line profile as a function of source
type is useful for future iron line surveys, since it is still unclear whether NLS1s are different from BLS1s in
terms of iron \ka emission.

Here we make a systematic comparison of  \ga~ and the iron \ka
emission lines of NLS1s and BLS1s based on \xmm observations of
X-ray luminous Seyfert 1 galaxies. Throughout this letter, we use
the cosmological parameters of \h0 , $\Omega_{m}=0.27$,
$\Omega_{\Lambda}=0.73$ (Komatsu et al. 2009).

\section{Sample and data reduction}
We compile a flux-limited ($f_{\rm 2-10\ keV} \geq 1 \times
10^{-12}$ erg s$^{-1}$ cm$^{-2}$ ) and radio-quite Seyfert 1 sample
from the \xmm archive, as listed in Table 1; 86 out of these 114
AGNs are included in the CAIXA catalogue (Bianchi et al. 2009a). The
redshift range of the present sample is $z<0.37$, with only seven
objects having $z
> 0.2$. The sample includes most of well-studied Seyfert 1 galaxies.

The data have been reduced as homogeneously as possible. Only the EPIC PN (Str\"uder et al. 2001) data are used
in our analysis. The SAS v7.0 software\footnote{{\it http://xmm.esac.esa.int/}}, with the corresponding
calibration files, are used for the data reduction.
  The X-ray events corresponding to patterns $0-4$ for the PN data are
  selected; hot or bad pixels are also removed. We extract the source
  spectra from a circle within $40''$ of the detected source position, with the
background being taken from the circular source-free regions; the CCD chip gaps are all avoided. The presence of
background flaring in the observation has been checked and removed using a Good Time Interval file. We also
check the pile-ups in the data with the SAS task {\it epatplot}. The response files are generated with the SAS
tools {\it rmfgen} and {\it arfgen}.

The spectral fittings are performed via a basic model expressed as
$phabs*zphabs*zedge (powerlaw + Nzgauss)$ in {\sc xspec} (Arnaud
1996) over the $2-10$ keV band; however the $2-7$ keV band is used
for 1H0707-495 since there is a sharp drop around 7 keV found in
this object (Boller et al. 2002). Errors are quoted at the 90\%
confidence level. For 108 out of 114 objects in Table 1, there exist
published spectral fitting results (see online material). We add an
additional Gaussian in our fits, if there was evidence for such a
component in the published fits; we only add a single Gaussian to
obtain line measurements in the remaining 6 AGNs. We note that we
have not tested independently the significance of including this
additional component.  All the fittings include the absorption due
to the line-of-sight Galactic column density (Dickey \& Lockman
1990). The neutral reflection model (Magdziarz \& Zdziarski 1995) is
also tried for each spectrum. However, these fits generally do not
improve significantly in the $2-10$ keV band compared with the
power-law fits. Thus we use the results returned from the basic
model.

\section{Results}
Panel (a) in Figure 1 shows the distribution of 2$-$10\ keV luminosity for the whole sample (thin solid line),
compared with that of NLS1s (thick solid line) and that of BLS1s (dotted line). The luminosity of the NLS1s and
BLS1s spans the same range from 10$^{41}$ to 10$^{45}$ \ergs; this is natural since this sample is flux-limited.

Panel (b) in Figure 1 shows the distribution of \ga~  for the whole
sample, compared with that of NLS1s and that of BLS1s. The vertical
dashed lines denote the weighted means of that of NLS1s, BLS1s and
the total sample, which are $2.04\pm0.04$, $1.74\pm0.02$,
$1.84\pm0.02$, respectively. We perform the generalized Wilcoxon
test constructed in ASURV \footnote{ {\it
http://astrostatistics.psu.edu/statcodes/asurv}} (Isobe et al. 1986)
to determine the probability that NLS1s and BLS1s in this sample
were drawn from the same parent population. We obtain the test
statistic value of 7.3, indicating that the probability $<0.01\%$.
This robust and non-parametric test shows compelling evidence for
the different photon indices between NLS1s and BLS1s in this sample.

Panel (c) in Figure 1 shows the distribution of equivalent width
(EW) of 6.4 keV NIKAL for the whole sample, compared with that of
NLS1s and that of BLS1s. The vertical dashed lines denote the
weighted means of that of NLS1s, BLS1s and the total sample, which
are $40\pm5$, $125\pm7$, $105\pm7$ eV, respectively. 34 out of these
114 AGNs have only upper limit data, with 20 out of 37 NLS1s and 14
out of 77 BLS1s, respectively; the values of EW are taken as half of
the upper limits. It can be seen that NIKALs from NLS1s are
systematically weaker than that from BLS1s. The generalized Wilcoxon
test shows the statistical value of 4.0, giving a probability of
$<0.01\%$ that NLS1s and BLS1s were drawn from the same parent
population.

Figure 2 shows \ga~ against FWHM\hb for AGNs with available broad \hb measurements. The anti-correlation is
significant, with the Spearman's coefficient of $-0.59$; the Spearman's probability associated with this
coefficient is $<0.01\%$. A linear regression by applying the parametric expectation-maximization
 algorithm gives,

\begin{equation}
\Gamma_{\rm 2-10\ keV}=(2.20\pm0.05)- (1.0\pm0.01)\left(\frac{\rm
FWHMH\beta}{10^{4}~ {\rm km~s^{-1}}}\right) .
\end{equation}

This suggests \ga $ >2.0$ for FWHM\hb $< 2000$ \kms in this sample. We therefore choose to label the AGN with
\ga $> 2.0$ as X-ray luminous NLS1 type. This criterion is useful for understanding the intermediate Seyfert
galaxies (Sy 1.8 and 1.9) whose broad-line regions are lightly obscured, and for AGNs lacking the follow-up
optical spectroscopies found in extragalactic surveys.

A flattening is also likely around FWHM\hb$\sim$4000 \kms in Figure
2. We give a separate fit for AGNs with FWHM\hb$<4000~\kms$
(so-called Population A sources, Sulentic et al. 2008),

\begin{equation}
\Gamma_{\rm 2-10\ keV}=(2.45\pm0.07)- (2.0\pm0.02)\left(\frac{\rm
FWHMH\beta}{10^{4}~ {\rm km~s^{-1}}}\right).
\end{equation}

\section{Broadened iron \ka lines}
N07 performed an \xmm survey of BIKAL based on 37 observations of 26
luminous Seyfert galaxies. Since data with high ratios of S/N in the
hard X-ray band are required to characterize the line profiles in
detail, they only selected the objects with at least 30 thousands
net counts in their EPIC PN spectra in the $2-10$ keV band. They
derived the BIKAL parameters from a homogeneous spectral analysis
after taking into account the NIKAL emission and the absorption due
to a zone of ionized gas along the line of sight.

Here we re-analyse the results obtained in N07. Figure 3 plots the distribution of BIKAL EW of NLS1s, compared
with that of BLS1s listed in N07. Most of BLS1s show weak BIKAL emission. Plausibly NLS1s show a more
homogeneous distribution. However, the generalized Wilcoxon test shows the statistic value of 1.4, indicating a
probability of 0.15 that NLS1s and BLS1s were taken from the same parent population. This makes it difficult to
draw a firm conclusion. The intrinsic width ($\sigma$) of BIKAL denotes the velocity dispersion of the line
emission region. We also study $\sigma$ listed in N07 as a function of FWHM\hb in Figure 4. All the large values
($\sigma>0.5$ keV) correspond to FWHM\hb $ \leq 4000 ~\kms$.

\section{Discussion}
A NLS1 galaxy is conventionally defined from its optical line width
(FWHM\hb $< 2000$ \kms). Zhu et al. (2009) had shown that for
optically selected AGNs, NLS1s and BLS1s have the same average
accretion rate (\ll) (see Figure 17c in Zhu et al 2009); however,
the black hole mass is significantly correlated with the broad
emission line width (see Figure 17a in Zhu et al 2009). They arrived
at this conclusion by modeling the broad emissions lines of the SDSS
sample with a physically realistic two-component model. For the flux
limited (i.e., X-ray luminous) AGN sample used here, the luminosity
distribution of NLS1s is indistinguishable from that of BLS1s (see
Figure 1a). Therefore, NLS1s must have on the average higher \ll~
than BLS1s in this X-ray flux limited sample, given that NLS1s
contain less massive black holes. Consequently the X-ray luminous
sample presented here is biased towards a higher \ll~ for the
narrower H$\beta$, unlike a significant fraction of the local
optical AGN population (Heckman et al. 2005). Thus the
anti-correlation between \ga~ and FWHM\hb shown in Figure 2 must be
due to the higher \ll~ for the narrower FWHM\hb, in agreement with
previous works on the correlation between \ga~ and \ll ~(e.g., Wang
et al. 2004). This is further supported by the mean \ll~ of
$\sim1.1$ for NLS1s and 0.16 for BLS1s, respectively, as calculated
from the data in Bianchi et al (2009a; see Table 1). Vasudevan \&
Fabian (2007) found that the bolometric correction, $L_{\rm
Bol}/L_{\rm 2-10\ keV}$, depends on \ll~, with a transitional region
at \ll $\sim0.1$; below which the bolometric correction is typically
15-25, and above which it is typically 40-70. Thus NLS1s, with the
mean \ll $\sim$ 1 in Table 1, have a higher bolometric correction,
resulting in a smaller ratio of X-ray luminosity to bolometric
luminosity than BLS1s. This also agrees with the conclusion of
Bianchi et al. (2009b) that NLS1s are X-ray weaker, relative to
BLS1s.

NIKAL are almost ubiquitous features in the present sample. The weighted mean width of NIKAL is $1350\pm250$
\kms for a small \chandra sample (Yaqoob \& Padmanabhan 2004), more than a factor of 2 lower than the weighted
mean width of \hb of 3200$\pm$ 60 \kms (Nandra 2006). This provides compelling evidence that NIKAL is mainly
emitted from the inner boundary region of the dusty torus in an AGN (Zhu et al. 2009). This agrees with previous
suggestions that NIKAL unlikely originates from either the outer region of the accretion disk or the
self-gravity-dominated disk (Zhou \& Wang 2005) or the the broad-line region (Wu et al. 2009). In particular,
the X-ray variability can not account for the observed NIKAL EW variation (Jiang et al. 2007). That means, NIKAL
may give measures of the covering factor of the dusty torus, indicating the changing AGN populations (Zhou \&
Wang 2005; Zhu et al. 2009). The relatively weaker NIKAL EW in NLS1 can be explained in terms of models that the
radiation pressure blows the cold dusty gas away from the central engine (Fabian et al. 2008), or the outflow
scenario which can be related with the high \ll (Komossa et al. 2008). If this is true, some NLS1 galaxies may
have geometrically-thin tori, as indicated by their weaker NIKAL emission. Conversely, the tori in some AGNs can
be very geometrically-thick, as indicated by their very large NIKAL EW. NIKAL thus is very useful to identify
this rare type of buried AGNs, which are difficult to detect so far (Ueda et al. 2008).

BIKAL is believed to be associated with the accretion disk, probing the strong gravity of a black hole (Fabian
et al. 1989; Laor 1991). Recently, Brenneman \& Reynolds (2006) developed a disk-line model in which the black
hole spin is a free parameter. This makes it possible to constrain the space-time geometry close to distant
black holes via X-ray spectroscopy. The results from N07 show that most of BLS1 galaxies show weak BIKAL
emissions, with the EW less than 100 eV (Figure 3). This is in good agreement with the theoretical expectation
of Ballantyne (2009). The mean EW and $\sigma$ of BIKAL in NLS1s are larger than that of BLS1s, indicating a
smaller inner disk radius, in agreement with the expectation of the evaporation disk model, in which the inner
radius of the disk is anti-correlated with \ll(Liu et al. 1999).

\section{Conclusions}
We present an X-ray luminous ($f_{\rm 2-10\ keV} \geq 1 \times
10^{-12}$ erg s$^{-1}$ cm$^{-2}$) Seyfert 1 sample including 114
radio-quiet objects, with the $2-10$ keV luminosity ranging from
10$^{41}$ to 10$^{45}$ erg s$^{-1}$. The NLS1s and BLS1s span the
same luminosity range. The weighted mean of \ga~ of NLS1s, BLS1s and
the total sample are $2.04\pm0.04$, $1.74\pm0.02$, $1.84\pm0.02$,
respectively. The anti-correlation between \ga~ and FWHM\hb is
strong with a flattening at FWHM\hb$\sim4000~\kms$. We propose that
 \ga~$ > 2.0$ may be taken to indicate X-ray luminous
NLS1 type, reflecting the higher accretion rate, \ll~ in these
objects.

The observed ratio between the average line width of NIKAL and that of \hb provides evidence that NIKAL mainly
originates from the inner boundary region of the dusty torus in an AGN. The weighted means of the equivalent
width (EW) of NIKAL
 of NLS1s, BLS1s and the total sample are $40\pm5$, $125\pm7$,
$105\pm7$ eV, respectively. Other than a few cases, NIKAL from NLS1s are generally weaker than that from BLS1s,
indicating a smaller torus covering fraction in NLS1s. Some objects with an exceptionally larger EW, which may
indicate a larger covering factor of the dusty torus, may represent a rare type of buried AGNs. Based on our
re-analysis of the results in Nandra et al. (2007), plausibly the broadened iron K$\alpha$ lines from NLS1s show
a more homogenous distribution than that from BLS1s. All AGNs with large intrinsic widths of iron K$\alpha$
lines ($\sigma>0.5$ keV) have FWHM\hb $\leq 4000 ~\kms$. This may give important clues to target selections for
future iron line surveys.

\acknowledgments We acknowledge an anonymous referee for many
helpful comments to improve the manuscript.
 We thank useful discussions with Done, C., Li,
T.-P., Schulze, N., Lu, Y., Soria, R., Wang, J.-M., Ward, M., Zhang,
Y.-H. and Zhao, Y.-H. XLZ thanks the support from China postdoctoral
science foundation; SNZ thanks the support from Directional Research
Project of the CAS under project No. KJCX2-YW-T03, the National
Natural Science
 Foundation of China under grant Nos. 10821061, 10733010, 10725313,
 and 973 Program of China under grant 2009CB824800.



\clearpage

\begin{figure}
\centering
\includegraphics[width=5.5 cm, angle=270]{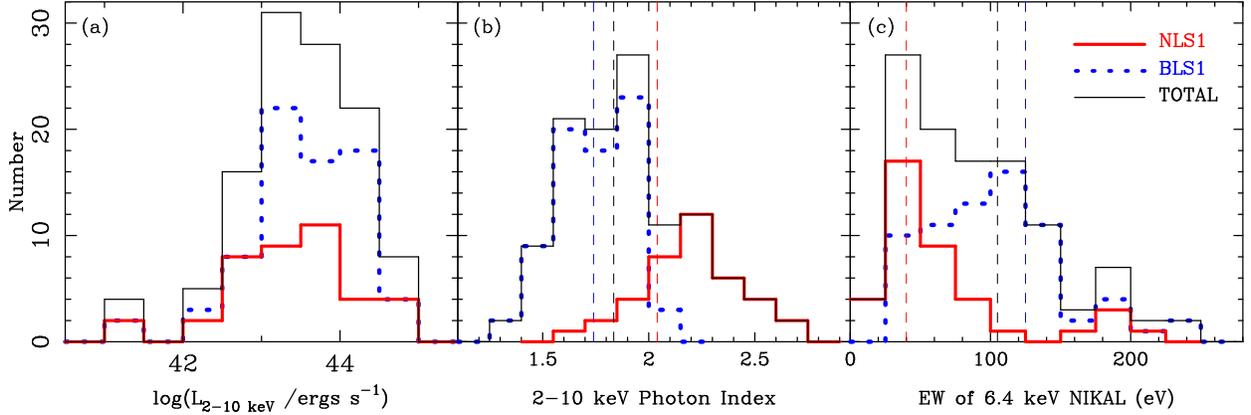}
\caption {(a): The distribution of $2-10$ keV luminosity for the whole sample (thin solid line), compared with
that of NLS1s (thick solid line) and that of BLS1s (dotted line); (b): The distribution of the hard X-ray photon
index for the whole sample, compared with that of NLS1s and that of BLS1s. The vertical dashed lines denote the
weighted means of that of NLS1s, BLS1s and the total sample at $2.04\pm0.04$, $1.74\pm0.02$, $1.84\pm0.02$,
respectively; (c):
 The distribution of the equivalent width of the 6.4 keV narrow iron
\ka lines for the whole sample, compared with that of NLS1s and that of BLS1s. The vertical dashed lines denote
the weighted means of that of NLS1s, BLS1s and the total sample at $40\pm5$, $125\pm7$, $105\pm7$ eV,
respectively.}
\end{figure}

\begin{figure}
\centering
\includegraphics[width=6.7 cm, angle=270]{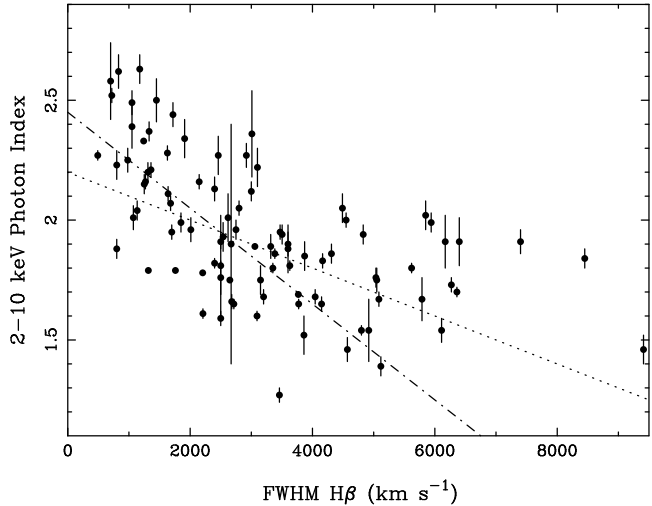}
\caption {The hard X-ray photon index against FWHM of the broad \hb
lines for AGNs with the \hb measurements available. The
anti-correlation appears strong, but a flattening is likely around
FWHM\hb$\sim$4000 \kms. The dotted line denotes the best fit for the
entire data given by Equation (1);  the dotted-dashed line denotes
the fit for AGNs with FWHM\hb $< 4000~ \kms$ given by Equation (2).}
\end{figure}

\begin{figure}
\centering
\includegraphics[width=6.7 cm, angle=270]{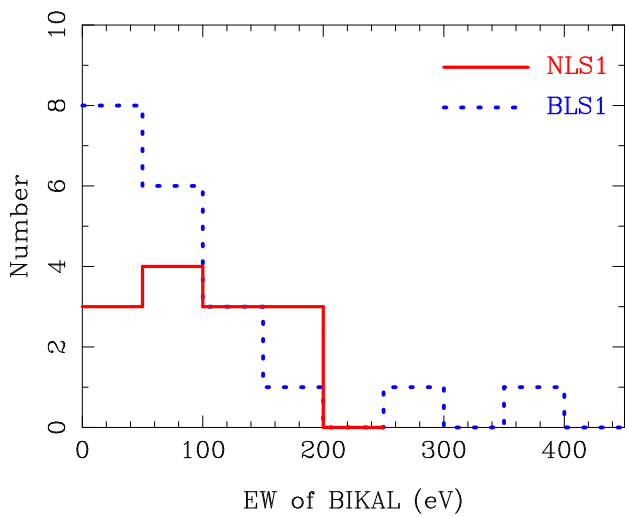}
\caption {The distribution of the equivalent width of the broadened iron \ka lines for NLS1s (thick solid line),
compared with that of BLS1s (dotted line) in Nandra et al. (2007). Plausibly NLS1s show a more homogeneous
distribution. }
\end{figure}

\begin{figure}
\centering
\includegraphics[width=6.7 cm, angle=270]{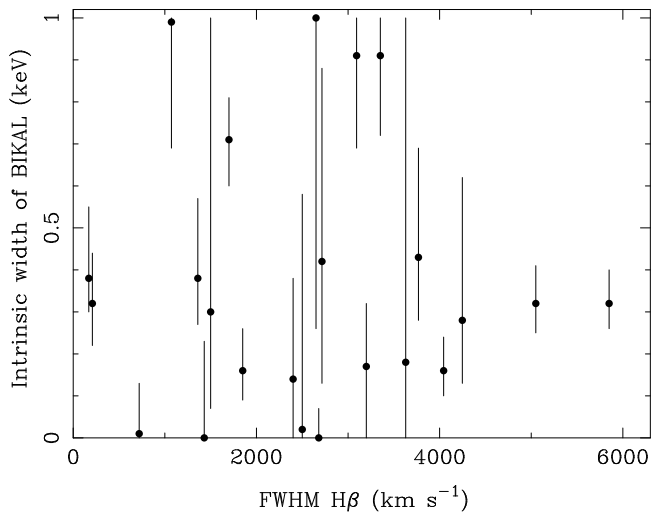}
\caption {The intrinsic width of broadened iron \ka lines as a
function of FWHM\hb in Nandra et al. (2007). All the large width
values (intrinsic width $\sigma>0.5$ keV) correspond to FWHM\hb $
\leq 4000 ~\kms$ }
\end{figure}

\begin{deluxetable}{llcclccr}
 \tabletypesize{\scriptsize} \tablecaption{ {\it
XMM-Newton} flux-limited
 sample of Seyfert 1 galaxies \label{tbl-2}} \tablewidth{0pt}
\tablehead{ \colhead{Source} & \colhead{Type} & \colhead{$f_{\rm
2-10\ keV}$} & \colhead{$\Gamma_{\rm 2-10\ keV}$} &
\colhead{EW(NIKAL)} &  \colhead{$L_{\rm 2-10\ keV}$}
&\colhead{FWHM(H$\beta$) } &
\colhead{log($L/L_{\rm Edd}$)}\\
\colhead{(1) } & \colhead{(2) } & \colhead{ (3) } & \colhead{(4) } &
\colhead{(5) } &\colhead{(6) } & \colhead{(7)} & \colhead{(8)}}

\startdata
Mrk 335     & NLS1 & 13  & $2.28\pm0.03$  & $\lesssim$ 54    & 43.27 & 1629 &0.05  \\
1 Zw 1      & NLS1 &7.9  & $2.33\pm0.01$  & $34(4-64)$        & 43.85 & 1240 & 0.12  \\
PG 0052+251 & BLS1 &6.3  & $1.83\pm0.03$  & $\lesssim70$      & 44.61 & 4165 &$-0.83$  \\
Q 0056-363  & BLS1 &2.3  & $2.00\pm0.03$  & $153(115-195)$    & 44.23 & 4550 & $-0.75$ \\
Ton S180    & NLS1 &4.5  & $2.25\pm0.05$  &  $\lesssim$ 64    & 43.62 & 980  & 0.29 \\
ESO 113-G10 & NELG &2.7  & $1.91\pm0.04$  &  $54(36-85)$      & 42.58 & ...  & ...  \\
Mrk 1152    & BLS1 &4.6  & $1.60\pm0.05$  &  $141(100-180)$   & 43.47 & ...  & ... \\
ESO 244-G17 & BLS1 &3.1  & $1.89\pm0.05$  &  $142(128-156)$   & 42.57 & 3317$^{a}$ &...  \\
Fairall 9   & BLS1 &18   & $1.73\pm0.03$  &  $139(113-165)$   & 43.97 & 6270  & $-1.72$ \\
NGC 526A    & NELG &17   & $1.41\pm0.06$  &  $78(51-105)$     & 43.14 & ...   &...   \\
Mrk 359     & NLS1 & 5.2 &1.88 $\pm$ 0.04 &  $205(139-274)$   & 42.50 & 800   & $-0.60$  \\
Mrk 1014    & NLS1 & 0.95& $2.27\pm 0.08$ &  $\lesssim310$    & 43.85 & 2460  & $-0.11$ \\
Mrk 586     & NLS1 &1.7  & $2.39\pm0.09$  &  $\lesssim 103$   & 44.05 & 1050  & $0.53$ \\
Mrk 590     & BLS1 &4.7  & $1.66\pm0.03$  &  $240(220-260)$   & 42.86 & 2680  & $-0.16$ \\
Mrk 1044    & NLS1 &6.2  & $2.20\pm0.04$  &  $186(125-247)$   & 42.55 & 1310  & $0.02$  \\
ESO 416-G02 & NELG &4.5  & $1.63\pm0.05$  &  $\lesssim206$    & 43.58 & ...   & ...   \\
ESO 198-G24 & BLS1 &3.9  & $1.91\pm0.10$  &  $104(55-153)$    & 43.27 & 6400  & $-1.36$ \\
RBS 416     & NLS1 &1.3  & $2.07\pm0.03$  &  $\lesssim81$     & 43.20 & 1680  & ...   \\
Mrk 609     & NELG &1.4  & $1.64\pm0.04$  &  $\lesssim299$    & 42.56 & ...   & ...   \\
Fairall 1116& BLS1 &5.3  & $1.86\pm0.04$  & $122(82-162) $    & 43.63 & 4310  & ...  \\
1H 0419-577 & BLS1 &7.7  & $1.27\pm0.03$  &  $\lesssim82$     & 44.33 & 3460  & $-0.60$  \\
3C 120      & BLS1 &32   & $1.78\pm0.01$  &  $54(30-78)$      & 43.91 & 2205  & $-0.72$ \\
RBS 553     & NELG & 1.6 & $1.84\pm0.04$  & $\lesssim210$     & 43.30 & ...   & ...  \\
H 0439-272  & BLS1 & 5.6 & $1.91\pm0.03$  & $60(30-90)$       & 43.98 & 2500  & ...  \\
MCG-01-13-25& BLS1   & 7.8 & $1.76\pm0.04$  & $135(37-233)$   & 42.85 & 5037$^{a}$ & ...  \\
HE 0450-2958& NLS1 & 2.0 & $2.16\pm0.03$  & $\lesssim$60      & 44.61 & 1270   & ...  \\
Ark 120     & BLS1 & 38  & $2.02\pm0.06$  & $38(17-55)$       & 43.95 & 5850   & $-0.99$ \\
MCG-02-14-09& BLS1 & 3.9 & $1.85\pm0.06$  &  $165(78-255) $   & 42.82& 3871$^{a}$ &...  \\
MCG+8-11-11 & BLS1 & 44  & $1.81\pm0.02 $ & $88(72-104)$      & 43.60 & 3630 & $-0.78$  \\
H0557-385   & BLS1 & 42  &  $1.89\pm0.01$ &  $86(50-122)$     & 44.03 & 3058$^{a}$ &...  \\
PKS 0558-504& NLS1 & 7.1 & $2.15\pm0.04$  &  $\lesssim$15     & 44.55 & 1250  & ...  \\
PMN J0623-6436&BLS1& 3.6 & $1.96\pm0.04$  &  $114(33-197)$    & 44.19 & 2750$^{a}$ & ...   \\
Mrk 6       & BLS1 & 22  & $1.81\pm0.21 $ & $95(71-109)$      & 43.21  & 2500 & ... \\
1H 0707-495 & NLS1 & 1.2 & $2.49\pm0.05$  & $59(33-98)$       & 42.67 & 1050  & $-0.04$  \\
UGC 3973    &BLS1  &  12 & $1.67\pm0.03$  & $178(160-196)$    & 43.12 & 5086  & $-1.25$  \\
ESO 209-G12 &BLS1  & 8.7 & $1.65\pm0.02$  & $104(60-151)$     & 43.51 & 3774$^{a}$ & ...  \\
PG 0804+761 & BLS1 & 11  & $1.96\pm0.05$  & $100(40-170)$     & 44.46 & 2012  & $-1.07$    \\
Fairall 1146& BLS1 & 9.5 & $1.57\pm0.02$  & $57(17-97)$       & 43.32 & ...   & ...   \\
PG 0844+349 & BLS1 & 5.5 & $2.16\pm0.03$  & $\lesssim$ 100    & 43.74 & 2148  & $-0.77$  \\
MCG+04-22-42& BLS1 & 16  & $1.86\pm0.02$  & $115(62-171)$     & 43.58 &  ...  & ... \\
Mrk 110     & NLS1 & 29  &  $1.79\pm0.01$ & $49(39-59)$       & 43.92 &  1760 & $-0.36$ \\
NGC 2992    & NELG & 85  &  $1.53\pm0.01$ & $56(46-66)$       & 42.97 & ...   & ...  \\
MCG-5-23-16 & NELG & 74  &  $1.90\pm0.03$ & $41(30$-$52)$     & 43.02 & ...   & ...  \\
PG 0947+396 & BLS1 & 1.9 &  $1.94\pm0.04$ & $120(60-180)$     &44.37 & 4830   & $-0.93$  \\
PG 0953+414 & NLS1 & 3.2 &  $2.12\pm0.04$ & $\lesssim $ 50    &44.73 & 3000   & $-0.05$  \\
HE 1029-1401& BLS1 & 11  &  $1.91\pm0.05$ & $\lesssim102$     &44.31 & 7400   & $-0.88$ \\
RE J1034+396& NLS1 & 0.9  & $2.58\pm0.16$ & $\lesssim171$     &42.57 & 700    & 0.16 \\
PG 1048+342 & BLS1 & 1.4    & $1.90\pm0.05$& $102(43-161)$    & 44.04 & 3600  & $-0.85$  \\
Mrk 728     & NELG & 4.0   & $1.68\pm0.05$ & $145(100-190)$   & 43.06 & ...   & ...  \\
NGC 3516    & BLS1 & 18  &  $1.80\pm0.02$  & $196(174-218)$   & 42.39 & 3353  & $-1.89$ \\
PG 1114+445 & BLS1 & 2.6 & $1.46\pm0.05$   & $100(65-135)$    & 44.16  & 4570 & $-0.86$   \\
PG 1115+407 & NLS1 & 1.3   & $2.44\pm0.05$  & $\lesssim$ 100  & 43.93 &1720   & ... \\
PG 1116+215 & NLS1 & 3.5  & $2.27\pm0.05$  & $\lesssim$ 80    & 44.49 & 2920  & $-0.01$  \\
RBS 980     & NELG & 2.4  & $1.79\pm0.05$  & $145(135-155)$       & 42.88 & ... & ...   \\
NGC 3783    & BLS1 &  59  &  $1.60\pm0.02$ & $116(103-129)$   & 43.03         &3093 & $-1.36$ \\
HE 1143-1810& BLS1 &  28  & $1.82\pm0.02$  & $53(39-65)$      & 43.82 & 2400 & ...  \\
NGC 4051    & NLS1 & 28   & $2.01 \pm0.05$   & $93(82-104)$   & 41.39 & 1072 & $-1.5$  \\
PG 1202+281 & BLS1 & 3.6  &$1.75\pm0.05$   & $\lesssim$ 80    & 44.43 & 5050 & $-0.45$  \\
NGC 4151    & BLS1 & 83   & $1.65\pm0.03$   & $187(184-190)$  & 42.22 & 4148 & $-1.39$ \\
PG 1211+143 & NLS1 & 3.2  & $1.79\pm0.01$  & $40(10-60)$      & 43.70 & 1317 & $-0.56$ \\
Mrk 766     & NLS1 &  45  & $2.21\pm0.03$  & $41(20-61)$      & 43.16 & 1360 & $-0.26$  \\
PG 1216+069 & BLS1 & 1.4  & $1.67\pm0.09$  & $\lesssim $70    & 44.72 & 5790 & $-1.50$  \\
Mrk 205     & BLS1 & 7.4  & $1.75\pm0.06$  & $60(35-85)$      & 43.95 & 3150 & $-0.57$  \\
Ark 374     & BLS1 & 3.2  & $1.94\pm0.04$  & $85(38-128)$     & 43.49 & 3504 & ...  \\
NGC 4579    & NELG & 3.9  & $1.78\pm0.08$  & $112(70-144)$    & 41.11 & ...  & ...  \\
NGC 4593    & BLS1 & 65  & $1.69\pm0.01$  &  $98(77-119)$     & 43.07 & 3769 & $-0.79$  \\
Was 61      & NLS1 & 4.9 & $2.11\pm0.03$  & $41(20-62)$       & 43.33 & 1640 & ...   \\
PG 1244+026 & NLS1 &  2.6 & $2.62\pm0.07$ & $\lesssim$ 146    & 43.15 & 830  & $0.12$ \\
ESO 323-G77 & BLS1 & 9.2  & $1.76\pm0.07$ & $111(79-144)$     & 42.67 & 2500 & ...  \\
PG 1307+085 & BLS1 & 1.8  & $1.52\pm0.08$ & $\lesssim$ 110    & 44.08 & 3860 & $-1.18$  \\
NGC 5033    & NELG & 3.1  & $1.94\pm0.03$ & $109(59-169)$     & 41.11  & ... &... \\
PG 1322+659 & NLS1 & 1.3 & $2.22\pm0.08$  & $180(70-290)$     & 44.02 & 3100 & $-0.24$  \\
MCG-6-30-15 & NLS1 & 73   & $1.95\pm 0.03$& $52(42-62)$       & 42.90  & 1700 & $-0.81$   \\
IRAS 13349+2438&NLS1& 2.2  & $2.05\pm0.03$ & $\lesssim84$     & 43.81   & 2800 &$-0.58$   \\
NGC 5273    & NELG & 6.7  &  $1.42\pm0.09$ & $233(168-296)$   & 41.43   & ... & ... \\
4U 1344-60  & BLS1 & 49  &  $1.54\pm 0.13$ &   ...            & 43.24  & 4920$^{a}$ &... \\
IC 4329A    & BLS1 & 160 & $1.80\pm0.02$   &  $44(39-49)$     & 43.96 &  5620 & $-0.83$  \\
Mrk 279     & BLS1   & 16  &  $1.86\pm0.02$  & $78(58-101)$     &  43.50 & 3385 & $-0.81$ \\
PG 1352+183 & BLS1 & 2.1  &$1.88\pm0.10 $  & $150(70-230)$    & 44.13 &  3600   & $-0.30$  \\
PG 1402+261 & NLS1 & 1.9   & $2.34\pm0.08$ & $\lesssim$ 100   & 44.15 & 1910 & 0.26  \\
NGC 5506    & NLS1 & 85    & $1.99\pm0.04 $& $70(50-90)$      & 42.83 & 1850 & $-0.38$ \\
PG 1411+442 & BLS1 & 1.3   & $1.90\pm0.50$ & $225(135-315)$   & 43.40 & 2670 & $-1.38$  \\
PG 1415+451 & BLS1 & 1.2   &  $2.01\pm0.10$& $110(30-190)$    & 43.60 & 2620 & $-0.77$ \\
NGC 5548    & BLS1 &  38   & $1.68\pm0.03$ &  $62(53-71)$     & 43.39 & 4044 & $-1.63$  \\
PG 1416-129 & BLS1 &  1.7  & $1.54\pm0.05$ &  $60(30-90)$     & 43.88 & 6110 & $-1.45$   \\
PG 1425+267 & BLS1 &  1.9  & $1.46\pm0.06$ &  $200(60-340)$   & 44.94 & 9410 & ...   \\
Mrk 1383    & BLS1 &  6.8  & $1.99\pm0.04 $&  $141(100-182)$  & 44.10 & 5940 & $-1.07$  \\
PG 1427+480 & BLS1 & 1.1 & $ 1.93\pm0.06$  &  $90(40-140)$    & 44.20 & 2540 & $-0.41$   \\
PG 1440+356 & NLS1 & 3.7   & $2.50\pm0.09$ & $\lesssim$ 80    & 43.76 &  1450 & $0.09$  \\
PG 1448+273 & NLS1 & 1.9   & $2.37\pm0.04$  & $74(15-134)$    & 43.29 & 1330 & 0.43  \\
Mrk 841     & BLS1 & 15    & $1.95\pm0.03$  & $125(75-148)$   & 43.89 & 3470 & $-0.36$  \\
PKS 1514+00 & NELG & 1.7   & $1.70\pm0.05$ &  $\lesssim81$    & 43.05 &  ... & ...  \\
Mrk 290     & BLS1 & 9.3   & $1.59\pm0.03$ & $43(12-74)$      & 43.25 & 2500 & ...   \\
Mrk 493     & NLS1 & 3.6   & $2.23\pm0.06$&  $\lesssim 105$   & 43.22  & 800 & 0.21  \\
Mrk 876     & BLS1 &  3.5  & $1.84\pm0.04$  & $96(37-155)$    & 44.19 & 8450 & $-1.25$  \\
PG 1626+554 & BLS1  &  3.1  & $2.05\pm0.06 $ & $\lesssim$ 160  & 44.16 & 4490 & $-0.66$   \\
IRAS 17020+4544&NLS1& 5.8  & $2.27\pm0.02$  & $\lesssim 46$   & 43.70 &  490  & 0.51  \\
PDS 456     & NLS1 & 6.1   & $2.36\pm0.18$ & $\lesssim$ 18    & 44.77 &  3010 & $-0.11$ \\
IGR J17418-1212 & BLS1 & 13   & $1.93\pm0.04$  & $54(27-80)$  & 43.62 & ...   & ...  \\
Mrk 509     & BLS1 & 179   & $1.65\pm0.02$  &  $85(28-142)$    & 44.68 & 2715 & $-1.1$  \\
Mrk 896     & NLS1 & 3.2   & $2.04\pm0.04 $ &  $180(93-267)$   & 42.70 & 1135 & 0.17  \\
Mrk 1513    & NLS1 & 3.5   & $1.61\pm0.02$  & $64(29-101)$     & 43.51 & 2210 & ...  \\
CTS A08.12  & BLS1 & 7.3  & $1.39\pm0.04$ & $100(80-120)$      & 43.16 & 5118$^a$ & ...  \\
NGC 7213    & BLS1 &  34  & $1.68\pm0.03$  & $82(66-98)$       & 42.27 &  3200 & $-1.79$  \\
Mrk 304     & BLS1 & 6.4   & $1.91\pm0.11$  & $\lesssim$ 115   & 43.82 & 6170  & $-0.59$   \\
II Zw 177   & NLS1 & 0.95   & $2.63\pm0.06$  & $\lesssim$ 88   & 43.22 & 1176  & ... \\
NGC 7314    & NELG & 54   & $2.19\pm0.09$  & 42(22$-$62)       & 42.28 & ...   & ... \\
ESO 602-G31 & NELG & 5.0  & $1.77\pm0.04$  & 140(138$-$152)    & 43.10  &  ... &...     \\
UGC 12138   & NELG & 7.9   & $1.87\pm0.05$  & 124(114$-$134)   & 43.05  & ...  & ... \\
Ark 564     & NLS1 & 24  & $2.52\pm 0.03$ & $40(20-60)$        & 43.50 & 720  & 0.07  \\
MR 2251-178 & BLS1 &  29  & $1.54\pm0.02$  &  $\lesssim$ 74    & 44.46 & 4800 & ... \\
NGC 7469    & BLS1 &  26   & $1.75\pm0.01$  & $105(80-130)$    & 43.17 & 2650 & $-0.49$  \\
Mrk 926     & BLS1 &  30   & $1.70\pm0.02$  & $51(30-72)$      & 44.17 & 6360 & $-1.78$  \\
AM 2354-304 & NLS1 & 3.4   & $2.13\pm0.05$  & $\lesssim167$    & 42.85 & 2400 & ... \\
\enddata
\tablecomments{Col. (1): source name; Col. (2): source type. NELG:
Narrow-emission-line galaxy, NLS1: Narrow line Seyfert 1 galaxy,
BLS1: Broad line Seyfert 1 galaxy. NELGs are intermediate Seyfert
galaxies whose broad line regions are lightly obscured. We take
NELGs with $\Gamma_{\rm 2-10\ keV}> 2.0$ as NLS1s in our analysis;
Col. (3): absorption-corrected 2$-$10 keV flux, in unit of
10$^{-12}$ erg s$^{-1}$ cm$^{-2}$; Col. (4): 2$-$10 keV photon
index; Col. (5): equivalent width (EW) of the 6.4 keV narrow iron
K$\alpha$ line, in unit of eV. The intrinsic width is fixed at 10 eV
for the EW measurements; Col. (6): log of the 2$-$10 keV luminosity,
calculated from the 2$-$10 keV flux ; Col. (7): FWHM of the broad
\hb line, in unit of \kms. Col. (8): Eddington ratio, calculated
from the data available in Bianchi et al. (2009)a. $^{a}$ Object
with the FWHM\hb calculated from the FWHMH$\alpha$ using the
relation ${\rm FWHMH}\beta =1070 \times ({\rm FWHMH}\alpha
/1000)^{1.03}$ \kms given in Greene \& Ho (2005). These objects have
the FWHMH$\alpha$ measurements but lacking the FWHMH$\beta$
measurements. FWHMH$\alpha$ of these objects are taken from Pietsch
et al. (1998), Rodr\'iguez-Ardila et al. (2000) and Piconcelli et al
(2006), respectively. }
\end{deluxetable}

\end{document}